\documentclass[12pt,onesided]{article}

\usepackage[leqno]{amsmath}
\usepackage{amsthm}
\usepackage{amssymb}
\usepackage{amsfonts}
\usepackage{array}
\usepackage{hyperref}
\usepackage{rotating}
\usepackage{color}
\usepackage{verbatim} 

\usepackage{xy}
\input xy
\xyoption{all}

\def\bi{\begin{itemize}}
\def\ei{\end{itemize}}

\begin{document}

\title{A primer on reflexivity and price dynamics\\ under systemic risk}
\author{Tom Fischer\thanks{Institute of Mathematics, University of Wuerzburg, 
Emil-Fischer-Strasse 30, 97074 Wuerzburg, Germany.
Tel.: +49 931 3188911.
E-mail: {\tt tom.fischer@uni-wuerzburg.de}.
}\\
{University of Wuerzburg}} 
\date{This version: \today\\ First version: January 9, 2013}
\maketitle

\begin{abstract}
A simple quantitative example of a reflexive feedback process and the resulting price dynamics after an exogenous price shock
to a financial network is presented. Furthermore, an outline of a theory that connects financial reflexivity, which stems from
cross-ownership and delayed or incomplete information, and no-arbitrage pricing theory under systemic risk is provided.
\end{abstract}

\noindent{\bf Key words:} 
Asset valuation, counterparty risk, credit risk, cross-holdings, cross-ownership,
derivatives pricing, financial contagion, financial networks, no-arbitrage pricing, ownership structure, 
price dynamics, price shocks, reciprocal ownership, reflexive feedback process, reflexivity, structural model, systemic risk.\\

\noindent{\bf JEL Classification:} B41, C69, G12, G13, G32, G33, G38

\noindent{\bf MSC2010:} 91B24, 91B25, 91B52, 91G20, 91G40, 91G50


\section{Introduction}

Despite of George Soros's fame as one of the most successful speculators of the 20th century, his theory of reflexivity in finance -- first published
in Soros (1987) -- has not gained a lot of academic attention. A few notable exceptions are the works of Cross and Strachan (1997),  Bryant (2002), Kwong (2008) and Palatella (2010). 
While the latter two papers are concerned with mathematical models of reflexivity (both in terms of discrete
time dynamical systems) and while the article by Palatella indeed is on pricing models, it can certainly be said 
that aspects of reflexivity in {\em quantitative} finance have hardly been considered in the past. 
For instance, there seem to be no quantitative
models of reflexivity in finance which take into account systemic risk (stemming from the cross-ownership of financial assets) as a
major influence of valuation or price dynamics. This short note therefore has the intention to first explore an
extremely simple example of reflexivity and systemic risk in finance as the main cause of price dynamics in
the case of an exogenous price shock to an underlying asset. The next section will outline the mentioned numeric example
while recapitulating some of the basic ideas of reflexivity in finance according to Soros (1987, 2010).
The third section then relates the example to the no-arbitrage pricing theory under systemic risk which has
been developed in recent years (see Fischer (2012) and the references therein). It gives an outlook at a much more 
general model and a (for a lack of better terms) unifying theory which could have the potential to merge the existing theory
of risk-neutral valuation under systemic risk and a quantitative theory of reflexivity in finance. This connection might astonish
readers who know Soros's work, as he seems to view reflexivity and no-arbitrage theory (as an equilibrium theory) as somewhat irreconcilable (cf.~Soros (2010)). 

This short note might at a later stage be replaced with an article of the same or a similar title which 
would elaborate in more detail on the ideas presented in here.


\section{The example}

\label{The example}

Consider two financial entities called firm 1 and firm 2, these could be banks, hedge funds or similar, who regularly publish their quarterly results. We add systemic risk, or counterparty risk, into this small system by assuming that firm 1 owns 50\% of the equity of firm 2, and vice versa. Firm 1 holds commodities besides its share of firm 2. Firm 2 holds cash besides its share of firm 1. Both firms are levered through debt. Furthermore, we assume that the two firms are in a state of incomplete or delayed information in the sense that they base their quarterly published balance statements on the latest available published financial data which is one quarter back in the case of the respective other company's equity. Furthermore, and aside from up-to-date commodities prices in the case of firm 1, this (price) information is the only information the firms use in calculating their balance sheets.

How does reflexivity enter the picture? We have as cognitive functions (cf.~Soros (1987, 2010)) that each firm observes the other firm, i.e.~each
firm observes the publication of financial statements of the respective other firm. We have as manipulative functions (cf.~Soros (1987, 2010)) that each firm states their own results based on (incomplete or delayed) information about the other firm obtained through the cognitive function.
Note that in this simple example the cognitive and the manipulative function are imperfect in the sense that they rely on delayed 
information about the other firm's financial status. 

How does the reflexive feedback process (or loop) work (cf.~Soros (2010))? Assume that, at time 0, the two firms are in an equilibrium in the following sense: both firms have \$500 in debt, firm 1 owns \$1,000 worth of commodities, firm 2 owns \$1,000 cash. Let us assume that each company's equity is \$1,000. Then, as firm 1 and 2 each hold 50\% of the respective other firms equity (momentarily worth \$500), we obtain the following balance sheets (Assets = Liabilities + Equity) at time 0:
\begin{eqnarray}
\nonumber
\text{Firm 1:} \hspace{85mm} & & \\     
\nonumber
\$ 1,000 \text{ commodities} + \$ 500 \text{ share in Firm 2} &  = &  \$ 500 \text{ debt} +  \$ 1,000 \text{ equity},\\
\nonumber
\text{Firm 2:} \hspace{85mm} & & \\  
\nonumber
\$ 1,000 \text{ cash} + \$ 500 \text{ share in Firm 1}  & =  & \$ 500 \text{ debt} + \$ 1,000 \text{ equity}.                
\end{eqnarray}

Now, we assume an external shock: at time 1 (one quarter after time 0), the value of the commodities firm 1 holds has dropped from \$1,000 to \$700. The firms use their incomplete, delayed knowledge (cognitive function) to set up their new balance statements (manipulative function) at time 1:
\begin{eqnarray}
\nonumber
\text{Firm 1:} \hspace{85mm} & & \\     
\nonumber
\$ 700 \text{ commodities} + \$ 500 \text{ share in Firm 2} &  = &  \$ 500 \text{ debt} +  \$ 700 \text{ equity},\\
\nonumber
\text{Firm 2:} \hspace{85mm} & & \\  
\nonumber
\$ 1,000 \text{ cash} + \$ 500 \text{ share in Firm 1}  & =  & \$ 500 \text{ debt} + \$ 1,000 \text{ equity}.                
\end{eqnarray}
This new situation can be seen as a disequilibrium, as the balance sheet of firm 2 contains outdated information about firm 1 (it contains a 50\% share of firm 1's equity of \$1,000 from time 0, which is now only worth \$700, but firm 2 can not know this yet\footnote{For instance,
firm 2 might not know this, because the balance sheet of firm 1 does not state what commodities exactly it holds, so firm 2 can obtain the
value of the equity of firm 1 only as stated on that firm's balance sheet.}), and vice versa. A reflexive feedback process has been set in motion (see also
Soros (2010)), as next quarter (at time 2) firm 2 will recognize the changed situation and adapt. In turn, this will lower firm 2's equity, which will picked up by firm 1 another quarter later, and so on. A vicious circle, or downward spiral, has been started (cf.~Fischer (2012)), 
but, as we will see below,
a new equilibrium will be reached in the end. This situation would therefore have to be characterized as a {\em negative} feedback loop in
the sense of Soros (2010), which in the limit leads to a stable setup, rather than a {\em positive} feedback loop, which would lead to
boom-bust-like dynamics for asset prices (cf.~Soros, 2010).

\begin{figure}
\center
\hspace{0mm}\includegraphics[bb= 1 275 600 1000, scale=0.85, clip=TRUE, ]{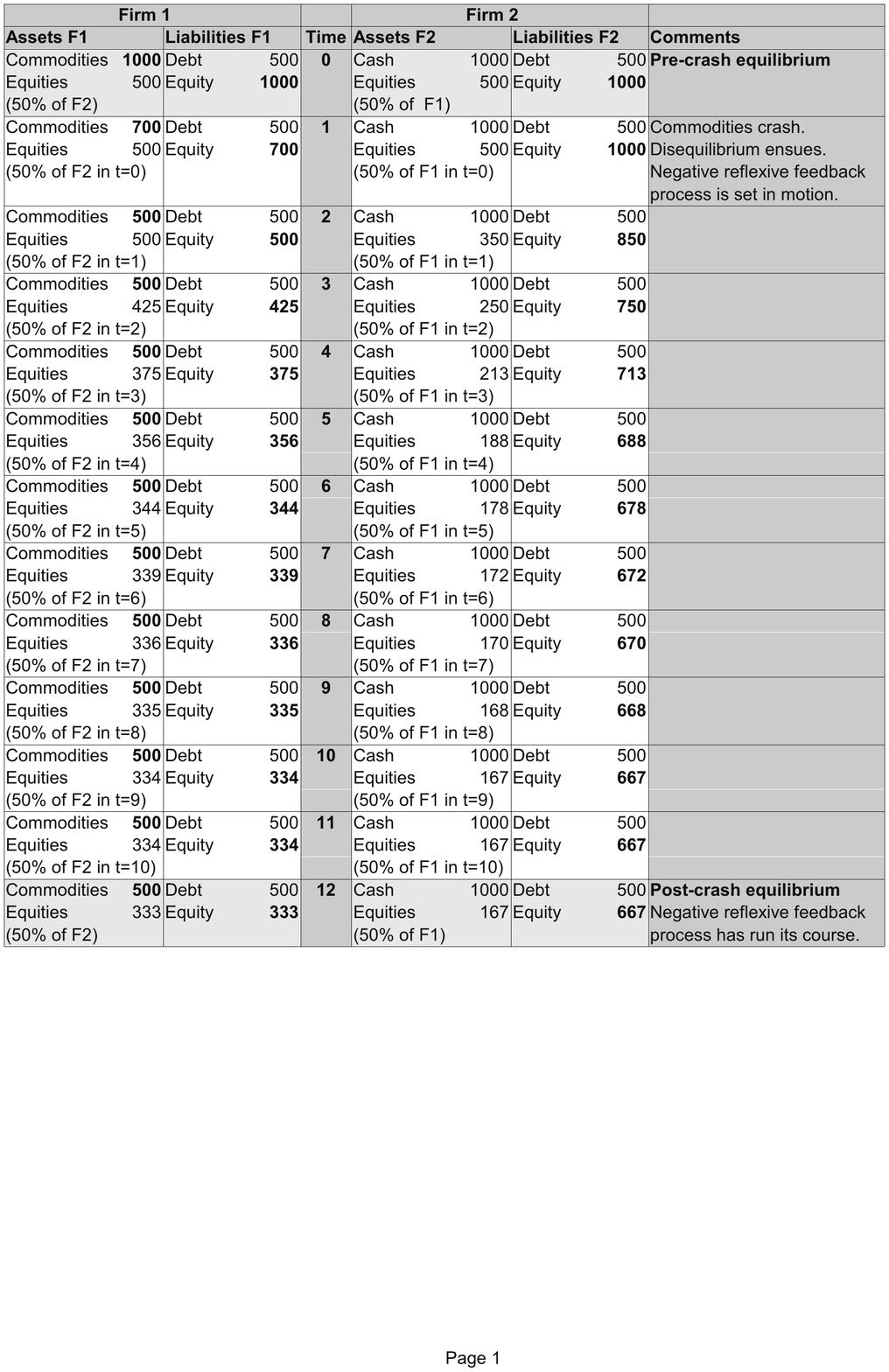}
\caption{\label{fig:example1} Example: detailed price information over time.}
\end{figure}

\begin{figure}
\center
\includegraphics[bb= 50 220 600 600, scale=0.9, clip=TRUE, ]{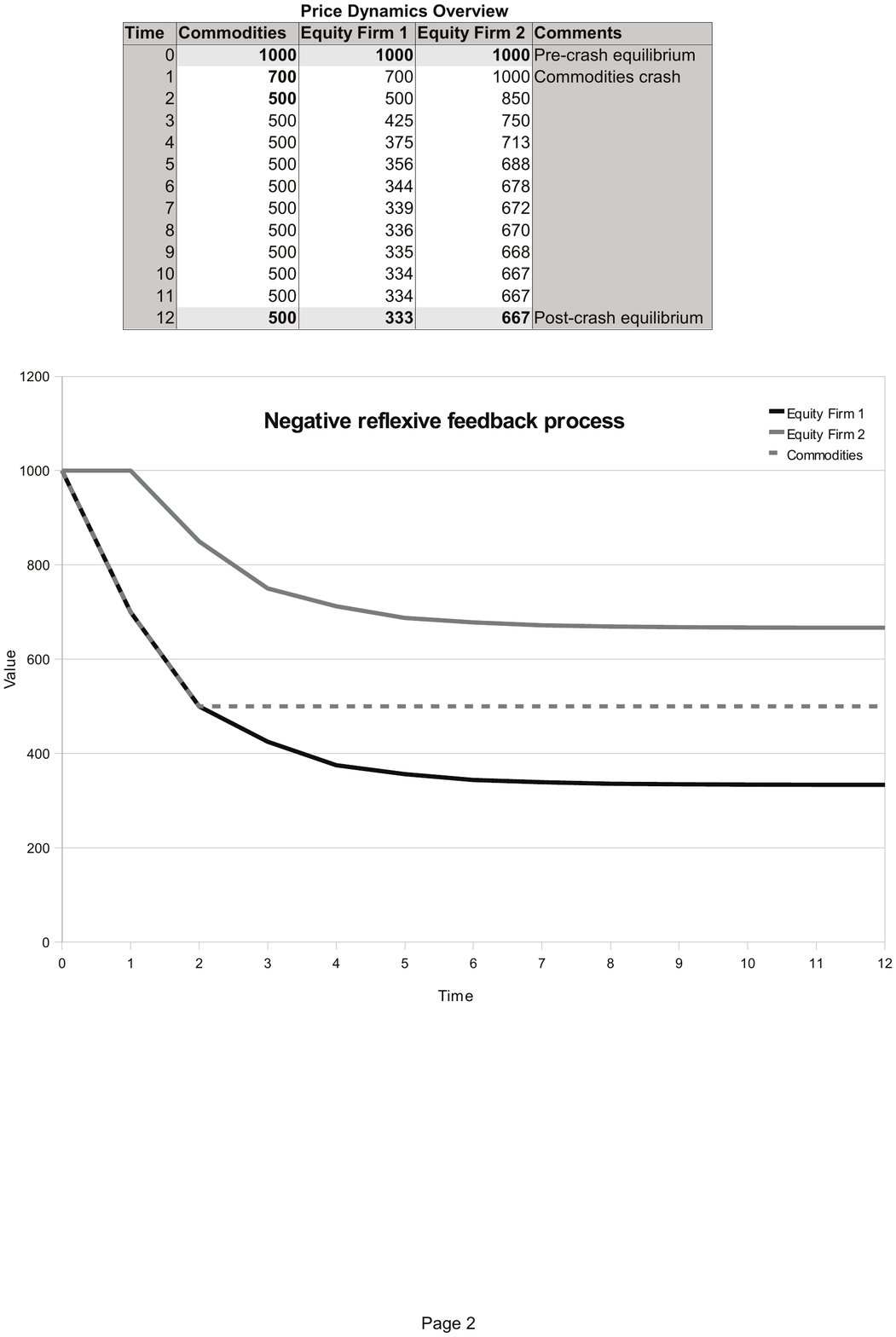}
\caption{\label{fig:example2} Chart of example price dynamics over time.}
\end{figure}

For the sake of this example, assume now that commodities correct one more time, to stabilize at \$500 at time 2. The new equilibrium, that would be reached after this (negative) reflexive feedback process has run its course, is:
\begin{eqnarray}
\nonumber
\text{Firm 1:} \hspace{85mm} & & \\     
\nonumber
\$ 500 \text{ commodities} + \$ 333.33 \text{ share in Firm 2} &  = &  \$ 500 \text{ debt} +  \$ 333.33 \text{ equity},\\
\nonumber
\text{Firm 2:} \hspace{85mm} & & \\  
\nonumber
\$ 1,000 \text{ cash} + \$ 166.67 \text{ share in Firm 1}  & =  & \$ 500 \text{ debt} + \$ 666.67 \text{ equity}.            
\end{eqnarray}
This new equilibrium is indeed only the limiting case of the reflexive feedback loop (compare Soros (2010)) and, mathematically, the equilibrium is never reached\footnote{For all practical purposes, however, it is reached after some time. See also Fig.~\ref{fig:example1} and Fig.~\ref{fig:example2}.}. 

Figures \ref{fig:example1}, \ref{fig:example2} and  \ref{fig:example3} provide a more detailed outline of the actual price dynamics
over time, as well as a chart, where the convergence to the limiting equilibrium can be seen. 


\section{Towards a more general theory}

It is clear that the extremely simple example of reflexivity in the previous section only captures one small aspect of reflexivity in finance
(in this case with respect to delayed or incomplete price information) as understood by Soros (1987, 2010).
However, the theory behind the presented example could be an important link between reflexivity and the existing 
standard no-arbitrage pricing theory of mathematical finance which, so far, could be termed as non-reflexive. 

What seems to be at hand with the outlined example is a {\em dynamic, reflexive systemic risk model under partial (or delayed) information}. 
However, if one compares the recursive procedure that was used in Section \ref{The example} to the methods used in Fischer (2012), 
it becomes immediately clear that
this procedure corresponds to the repeated iteration of the function $\Phi$ in Fischer (2012), that was used in a Picard Iteration
to determine the (in this case: unique) no-arbitrage prices of assets and liabilities in a system with systemic risk. Only, in Fischer (2012),
this recursive procedure was understood as a method to instantaneously price assets under full information
(an instantaneous equilibrium), while, in the reflexive example of Section \ref{The example}, the procedure needs to be 
re-interpreted in a quite different manner, namely as the convergence
over time (by means of a reflexive feedback process) to an equilibrium under incomplete information. 
Expressed differently, the recursion now describes the price dynamics over time during a market adjustment after an external price shock.
From Fischer (2012) and the references therein, it is clear that similar examples (that converge to a new equilibrium after an
external price shock in an underlying exogenous asset) can be constructed for arbitrarily many firms and much more complicated ownership structures and liabilities. For instance, derivatives and different seniorities of debt can be incorporated into this model. 

In this sense, the example of Section \ref{The example} appears to show a way to the development of more quantitative research
about  and more applications of reflexivity in finance in a manner that is consistent with existing no-arbitrage pricing theory. 
Note that while the mentioned iteration of a function $\Phi$ and convergence to a limiting equilibrium
in a time-discrete setup seems to imply a certain similarity of our ideas with those of Kwong (2008) and Palatella (2010), there are 
several important differences that fundamentally distinguish our ideas from these two (distinct) approaches to reflexivity in finance.
To mention one such difference, our approach is focussed on systemic risk stemming from the cross-ownership of liabilities in a
network of financial firms, which is a feature that Kwong (2008) and Palatella (2010) do not consider. Mathematical details would have to be outlined in
an extended version of this paper.

As a final, more philosophical remark regarding the ideas presented above, note that reflexivity as outlined here essentially means that the market (or a network of financial firms) constantly watches itself and then reacts to its observations. However, because of an information delay, this procedure is subject to incomplete (or flawed) information at any time it is in progress.



\begin{figure}
\center
\hspace{0mm}\includegraphics[bb= 25 600 600 1000, scale=0.9, clip=TRUE, ]{Table_2.pdf}
\caption{\label{fig:example3} Example overview.}
\end{figure}


\end{document}